\documentclass[
superscriptaddress,twocolumn,prl,amsfonts,showpacs]{revtex4}

\RequirePackage[utf8]{inputenc}
\RequirePackage[T1]{fontenc}

\RequirePackage[intlimits]{amsmath}
\RequirePackage{amssymb,amstext,color,xspace}

\providecommand{\lp}{\left(}
\providecommand{\rp}{\right)}
\providecommand{\lsqr}{\left[}
\providecommand{\rsqr}{\right]}
\providecommand{\lbr}{\left\{}
\providecommand{\rbr}{\right\}}

\providecommand\xspacedMathSymbol[1]{\ensuremath{{#1}}\xspace}

\providecommand\x{\xspacedMathSymbol{\mathbf{r}}}

\providecommand\p{\xspacedMathSymbol{\mathbf{p}}}
\providecommand\vv{\xspacedMathSymbol{\mathbf{v}}}
\providecommand\n{\xspacedMathSymbol{\mathbf{n}}}

\providecommand\DV{\xspacedMathSymbol{\mathbf{D}}}
\providecommand\EV{\xspacedMathSymbol{\mathbf{E}}}
\providecommand\PV{\xspacedMathSymbol{\mathbf{P}}}
\providecommand\AV{\xspacedMathSymbol{\mathbf{A}}}
\providecommand\PiV{\xspacedMathSymbol{\boldsymbol{\Pi}}}

\providecommand\Dcal{\xspacedMathSymbol{\mathcal{D}}}
\providecommand\Gcal{\xspacedMathSymbol{\mathcal{G}}}

\providecommand\Qfrak{\xspacedMathSymbol{\mathfrak{Q}}}
\providecommand\Rfrak{\xspacedMathSymbol{\mathfrak{R}}}

\providecommand\Ltwo[1][\null]{\xspacedMathSymbol{L^2_{#1}}}

\providecommand\epsNot{\xspacedMathSymbol{\varepsilon_0}}

\providecommand\intD[1][\null]{\int_\Dcal d^3r#1}

\providecommand{\surfaceD}{\xspacedMathSymbol{\partial\Dcal}}

\providecommand\modefunction[1][\lambda]{\xspacedMathSymbol{\boldsymbol{\varphi}_{#1}}}
\providecommand\modefunctionconj[1][\lambda]{\xspacedMathSymbol{\boldsymbol{\varphi}_{#1}^*}}

\providecommand{\CHgroup}{\xspacedMathSymbol{\mathbb{H}_2}}

\providecommand{\at}[2]{\left.#1\right\vert_#2}

\providecommand\rot[2][\null]{\nabla#1\times#2}
\providecommand\grad[2][\null]{\nabla#1#2}
\renewcommand\div[2][\null]{\nabla#1\cdot#2}

\providecommand{\ran}[1]{\xspacedMathSymbol{\text{ran}({#1})}}
\providecommand{\dom}[1]{\xspacedMathSymbol{\text{dom}({#1})}}
\renewcommand{\ker}[1]{\xspacedMathSymbol{\text{ker}({#1})}}

\providecommand{\threeSpace}{\xspacedMathSymbol{\mathbb{R}^3}}

\providecommand{\eref}[1]{{\nolinebreak[4] (\ref{#1})\xspace}}
\providecommand{\Eref}[1]{Eq.\eref{#1}}

\providecommand{\dv}{\xspacedMathSymbol{\boldsymbol{d}}}

\newcommand{\longwl}[1]{{#1}^<}

\begin{document}

\title{Elimination of the A-square problem from cavity QED}

\newcommand{\ourAddress}{\affiliation{Institute for Solid State Physics and Optics, Wigner Research Centre for Physics,\\Hungarian Academy of Sciences, P.O. Box 49, H-1525 Budapest, Hungary}}

\author{András Vukics}
\email{vukics.andras@wigner.mta.hu}
\ourAddress
\author{Tobias Grießer}
\affiliation{Institut für Theoretische Physik, Universität Innsbruck,\\Technikerstraße 25, 6020 Innsbruck, Austria}
\author{Peter Domokos}
\ourAddress

\begin{abstract}
\noindent We generalize the Power--Zineau--Woolley transformation to obtain a canonical Hamiltonian of cavity quantum electrodynamics for arbitrary geometry of boundaries. This Hamiltonian is free from the A-square term and the instantaneous Coulomb interaction between distinct atoms. The single-mode models of cavity QED (Dicke, Tavis--Cummings, Jaynes--Cummings) are justified by a term by term mapping to the proposed microscopic Hamiltonian. As one straightforward consequence, the basis of no-go argumentations concerning the Dicke phase transition with atoms in electromagnetic fields dissolves.
\end{abstract}


\maketitle


The fundamental  description of the interaction of atomistic matter with the electromagnetic field in the Coulomb gauge is known to suffer from the presence of an awkward term containing the square of the vector potential. In most of the practical cases, in the framework of a diluteness assumption for the atoms, this term can be neglected and the observable effects  are ultimately accounted for in terms of a simplified model, such as the Jaynes--Cummings one, for example. In typical quantum optical systems, such a phenomenological approach with properly adjusted parameters usually gives a satisfactory quantitative accuracy. However, there are situations in which even the qualitative behaviour of the system is questionable because of the confusion around this term. A prominent example is the Dicke model, where the very existence of the predicted superradiant phase transition depends on the validity of the adopted effective model.\cite{Rzazewski75,Knight78,Rzazewski91,Nataf10} Further discrepancies due to the A-square term occur in relation with novel artificial systems in which the electromagnetic field confined into a small volume is coupled to  some kind of polarizable material in the so-called ultrastrong coupling regime.\cite{Ciuti05,Todorov10}

In this Letter, we show that cavity quantum electrodynamics, i.e., when the field itself as well as the light-matter interaction are significantly influenced by the presence of boundaries, can be established at a fundamental level on a Hamiltonian which eliminates the problem of the A-square term. We present a canonical transformation which makes manifest that this term is compensated by a dipole-dipole interaction term, and the remaining terms are of a simple linear form.\cite{Keeling07} From our approach it follows, for example, that there is no principle that would prevent the superradiant phase transition in the case of an ensemble of atomic dipoles in a cavity. The canonical transformation is analogous to the Power--Zienau--Woolley (PZW) transformation in free space, however, in our approach we allow for  arbitrary geometry, thereby treating  general cavity QED system. All our vector fields are thus defined on a generic (possibly even multiply connected) domain \Dcal in the three-dimensional real space bounded by (possibly several disjunct) sufficiently smooth surfaces \surfaceD, which consist of a perfect conductor. Overall, \Dcal is assumed to be bounded.

Consider an arbitrary number of point charges coupled to the electromagnetic field confined into \Dcal. In the Coulomb (minimal-coupling) gauge, defined by
\begin{equation}
\label{eq:CoulombGaugeCondition}
\div{\AV}=0,
\end{equation}
the Hamiltonian of the system reads:
\begin{subequations}
\label{eq:CoulombGaugeContext}
\begin{equation}
\label{eq:HamiltonianCG}
H=\sum_\alpha\frac{\lsqr\p_\alpha-q_\alpha \AV(\x_\alpha)\rsqr^2}{2m_\alpha}+\frac\epsNot2\intD\,(\grad{U})^2+H_\text{field},
\end{equation}
with \(U\) being the scalar potential, \(\p_\alpha\) the canonical momentum of particle \(\alpha\) conjugate to its position \(\x_\alpha\), and
\begin{equation}
H_\text{field}=\frac\epsNot2\intD\lsqr\lp\frac\PiV\epsNot\rp^2+c^2\lp\rot{\AV}\rp^2\rsqr, 
\end{equation}
\end{subequations}
with \(\PiV=\epsNot\partial_t\AV\) being the momentum conjugate to \AV. 

An important observation is that, \emph{unlike in free space,} the condition \eref{eq:CoulombGaugeCondition} does not fix the potentials completely. The remaining freedom of choosing the potentials within the Coulomb gauge amounts to a freedom in choosing different constant values for \(U\) on each of the connected components of \(\surfaceD,\) which will result in various configurations of condensator fields carried by \(U\). Our choice here will be to set
\begin{equation}
\label{eq:boundaryCondition}
\at U\surfaceD=0\text{ and }\at{\AV\times\n}\surfaceD=0.
\end{equation}
Together with \Eref{eq:CoulombGaugeCondition}, the latter condition makes up for the vector potential satisfying both the electric and magnetic boundary conditions.\footnote{The freedom of choosing the potentials within the Coulomb gauge is equivalent also to a freedom of fixing how the inclusion of the cohomological fields introduced later in Eq. (\ref{eq:HH_decomposition}), is shared between the scalar or the vector potential. With our fixing of the potentials within the Coulomb gauge, what we attain is that \[U\in\dom{\text{grad}_0}\text{ and }\AV\in\ker{\text{div}_0},\] that is, the electrostatic and radiative parts of the dynamics take place in the two distinct orthogonal subspaces listed later in Eq. (\ref{eq:HH_decompositionBoundary}), the cohomological components of \EV (condensator fields) being attributed solely to \AV. Note that the form of the Hamiltonian (\ref{eq:HamiltonianCG}) depends on this decomposition result, since this ensures that there are separate electrostatic and radiative terms in the Hamiltonian, with no overlap between the two.}

The electric-dipole approximation to this Hamiltonian can be obtained in two steps. Step 1 (long-wavelength approximation): We assume that the individual point charges form (a certain number of) spatially separated, well-localized clusters, that is, atoms. Then, instead of \(\sum_\alpha\) there appears \(\sum_A\sum_{\alpha\in{A}}\). We neglect all radiative effects on the intra-atomic scale, that is, we set \(\AV(\x_\alpha)=\AV(\x_A),\) where \(\x_A\) is the position of that atom \(A\) which incorporates the charge \(\alpha\). Step 2: We assume that the atoms have only electric dipole moment, that is, no net charge and no further electric or magnetic moments.

Upon the first assumption, we split the Coulomb (electrostatic) term into intra- and inter-atomic parts, and take the intra-atomic part as identical to the one in free space, under the assumption that the distance of atoms from the boundary is much larger than the atomic radius. The electric-dipole order of the Hamiltonian in Coulomb gauge then reads:
\begin{subequations}
\label{eq:H_CG:ED}
\begin{multline}
H_\text{ED}=\sum_A\bigg[H_A-u\,\p_A\cdot\AV(\x_A)+v\,\AV^2(\x_A)\\+V_\text{Coulomb}^\text{dipole-self}(\x_A)+\sum_B V_\text{Coulomb}^\text{dipole-dipole}(\x_{A-B})\bigg]+H_\text{field},
\end{multline}
where \(u\) and \(v\) are constants composed of the \(m_\alpha\)s and \(q_\alpha\)s. The single-atom Hamiltonian reads
\begin{equation}
\label{eq:SingleAtomCG}
H_A=\sum_{\alpha\in{A}}\bigg(\frac{{\p_\alpha}^2}{2m_\alpha}+\frac{q_\alpha}{8\pi\varepsilon_0}\sum_{\underset{\beta\neq\alpha}{\beta\in A}}\frac{q_\beta}{|\x_\alpha-\x_\beta|}
\bigg).
\end{equation}
\end{subequations}

It is this Hamiltonian (\ref{eq:H_CG:ED}) that is usually taken as the starting point of cavity QED. However, it is fraught with the following problems: (i) the canonical momentum of the atoms does not equal their kinetical momentum; furthermore, as we mentioned, (ii) the presence of the A-square term, which yields creation and annihilation of pairs of photons; and finally, (iii) there appears an instantaneous electrostatic interaction between remote atoms (\(V_\text{Coulomb}^\text{dipole-dipole}\)) and an interaction of a single dipole with its own induced surface charges (\(V_\text{Coulomb}^\text{dipole-self}\)). The former is \emph{influenced,} while the latter is \emph{created} by the presence of the boundaries \cite{Vukics2012}.

In free space, these weaknesses can be dissolved by performing the PZW transformation on the minimal coupling Hamiltonian \eref{eq:HamiltonianCG} to the multipolar-coupling gauge (cf.~Ref.~\cite{CDG} Chapter IV.C). Here, inspired by the free-space procedure, we elevate this transformation onto a very general level, which allows for an arbitrary domain \Dcal and boundaries \surfaceD, i.e.~for a general cavity QED scenario.

The transformation that we adopt is canonical, defined by the Type-2 generating function
\begin{subequations}
\label{eq:canonicalTransformation}
\begin{equation}
\label{eq:generatingFunction}
G_2\equiv\intD\,\AV\cdot\lp\PiV'+\Rfrak\PV\rp+\sum_\alpha\x_\alpha\cdot\p'_\alpha,
\end{equation}
which yields a displacement of the momenta
\begin{align}
\PiV&=\frac{\delta{G_2}}{\delta\AV}=\PiV'+\Rfrak\PV,\\
\p_\alpha&=\frac{\partial{G_2}}{\partial\x_\alpha}=\p_\alpha'+\frac{\partial}{\partial\x_\alpha}\intD\,\AV\cdot\PV.
\end{align}
\end{subequations}
At this point, \PV is an arbitrary vector, and \Rfrak is part of an orthogonal projector decomposition of the identity,
\begin{equation}
\label{eq:completeness}
\Qfrak+\Rfrak=\mathbf{id}_{\Ltwo[0]}.
\end{equation}
where \Ltwo[0]  is the subspace of the Hilbert space \(\Ltwo(\Dcal,\threeSpace)\) of square-integrable vector fields such that the elements of \Ltwo[0] satisfy the boundary condition that they are normal to the boundaries:
\begin{equation}
\Ltwo[0](\Dcal,\threeSpace)\equiv\lbr\vv\in\Ltwo(\Dcal,\threeSpace)\middle\vert\at{\vv\times\n}{\surfaceD}=0\rbr,
\end{equation}
which is of course nothing else than the boundary condition on the electric field (and hence the vector potential) at a perfectly conducting surface. 

In order that the transformation \eref{eq:canonicalTransformation} be canonical, \Rfrak must be a projector onto the divergence-free subspace of \Ltwo[0]:
\begin{equation}
\label{eq:RfrakDefined}
\Rfrak : \Ltwo[0]\rightarrow\ker{\text{div}_0},
\end{equation}
because this ensures that \AV in \Eref{eq:generatingFunction} can be treated as unconstrained. Here, \(\text{div}_0\) (and \(\text{curl}_0\) below) are the divergence (and curl) operators over \Ltwo, with the domain restricted to \(\Ltwo[0]\). The notation ‘ker’ refers to the kernel of the operator, that is, the set of such vectors as are mapped onto zero by the operator. Hence, both the Coulomb-gauge and the boundary conditions on \AV can be expressed by the single condition that \(\Rfrak\AV=\AV\).

The crucial result for us to build upon here is the Helmholtz–Hodge decomposition of \Ltwo \cite{dautray,binz10}, which reads:
\begin{equation}
\label{eq:HH_decomposition}
\Ltwo(\Dcal,\threeSpace)=\rlap{$\underbrace{\phantom{\ran{\text{grad}_0}\oplus\CHgroup}}_{\displaystyle\ker{\text{curl}_0}}$}
\ran{\text{grad}_0}\oplus\overbrace{\CHgroup\oplus\,\ran{\text{curl}}}^{\displaystyle\ker{\text{div}}},
\end{equation}
where \(\text{grad}_0\) is the gradient operator over \(\Ltwo(\Dcal,\mathbb{R})\) with its domain restricted to such scalar fields \(v\) as vanish on the boundaries: \(\at{v}\surfaceD=0\). The notation ‘ran’ refers to the range of the operator. In free space, (\(\Dcal=\threeSpace\)) \(\ran{\text{grad}_0}=\ker{\text{curl}_0}\) (longitudinal fields) and \(\ran{\text{curl}}=\ker{\text{div}}\) (transverse fields) holds, and the direct sum of the two makes up for the whole \(\Ltwo(\threeSpace,\threeSpace)\). For general domains, however, the dimension of \CHgroup is non-zero. The elements of \CHgroup are called \emph{cohomological fields,} and, when the electric field is in question, also \emph{condensator fields}. On the basis of \Eref{eq:HH_decomposition}, we can assert that
\begin{equation}
\label{eq:HH_decompositionBoundary}
\Ltwo[0]=\ran{\text{grad}_0}\oplus\ker{\text{div}_0}.
\end{equation}
From this equation, together with \Eref{eq:RfrakDefined} it follows that in the decomposition of the identity in \Eref{eq:completeness}, the \Qfrak projector must be defined as
\begin{equation}
\label{eq:QfrakDefined}
\Qfrak : \Ltwo[0]\rightarrow\ran{\text{grad}_0},
\end{equation}

We recall that in free space \Qfrak~\footnote{%
Since the explicit form of \Qfrak is not needed for our derivation, we merely note that on the subspace of those \(\vv\in\Ltwo\) whose divergence exists, it can be written as
\[\lp\Qfrak\vv\rp(\x)\equiv-\grad{\intD[']\lp\div[']{\vv\lp\x'\rp}\rp\Gcal\lp\x,\x'\rp},\]
where \Gcal is the Dirichlet Green’s function of the problem:
\[\Delta\Gcal(\x,\x')\equiv\delta(\x-\x')\text{ within }\Dcal\text{, and }\at\Gcal{\surfaceD}=0.\]} and \Rfrak~\footnote{%
The explicit form of \Rfrak will not be used, so we merely note that it can be expressed with the full set of transverse modes (\ref{eq:transverseModes}) as
\[\Rfrak=\sum_\lambda\modefunction\otimes\modefunction.\]} project onto the longitudinal and transverse components of vector fields, respectively.

The transformed Hamiltonian reads:
\begin{multline}
\label{eq:HamiltonianPZW}
H'=\sum_\alpha\frac1{2m_\alpha}\lsqr\p'_\alpha+\frac\partial{\partial\x_\alpha}\intD\,\AV\cdot\PV-q_\alpha \AV(\x_\alpha)\rsqr^2\\+\frac\epsNot2\intD\,(\grad{U})^2\\+\frac\epsNot2\intD\lsqr\lp\frac{\PiV'+\Rfrak\PV}\epsNot\rp^2+c^2\lp\rot{\AV}\rp^2\rsqr.
\end{multline}
So far, we have not specified \PV. Since according to \Eref{eq:boundaryCondition} the scalar potential is an element of the domain of \(\text{grad}_0\), \Eref{eq:QfrakDefined} allows us to impose the condition on \PV that
\begin{equation}
\label{eq:conditionOnPV}
\epsNot\grad{U}=\Qfrak\PV.
\end{equation}
Hence, on account of \Eref{eq:completeness} the electrostatic term in the second line of \Eref{eq:HamiltonianPZW} and the term containing \(\PV^2\) in the third line combine to give \(\frac1{2\epsNot}\intD\,\PV^2\).

Condition \eref{eq:conditionOnPV} is \emph{equivalent} to \footnote{To prove the equivalence, we first prove (\ref{eq:conditionOnPV}) \(\Longrightarrow\) (\ref{eq:polarizationField}):\[-\rho=\epsNot\Delta{U}=\div{\Qfrak\PV}=\div{(\Qfrak+\Rfrak)\PV}=\div\PV,\] where the first equality is the Poisson equation, the second is obtained by applying the \(\nabla\) operator on both sides of Eq. (\ref{eq:conditionOnPV}), the third is on account of \(\div{\Rfrak\PV}=0\), while the fourth reflects Eq. (\ref{eq:completeness}). To prove (\ref{eq:polarizationField}) \(\Longrightarrow\) (\ref{eq:conditionOnPV}) we proceed as \[0=\div{\lp\epsNot\grad{U}-\PV\rp}=\div{\lp\epsNot\grad{U}-\Qfrak\PV\rp},\]where the first equality follows from Eq. (\ref{eq:polarizationField}) and the Poisson equation, while in the second we applied again \(\div{\Rfrak\PV}=0\). It follows that the vector in parenthesis on the right-hand side is both in \(\ran{\text{grad}_0}=\ran\Qfrak,\) and \(\ker{\text{div}_0}\), which, on account of Eq. (\ref{eq:HH_decompositionBoundary}), cannot be true but for the zero vector, so that \(\epsNot\grad{U}=\Qfrak\PV\) must hold.}
\begin{equation}
\label{eq:polarizationField}
\div{\PV}=-\rho,
\end{equation}
which motivates us to identify the vector field \PV, so far introduced on purely mathematical grounds, with the physical notion of the polarization density.

Besides the condition \eref{eq:conditionOnPV}, the following condition on the other orthogonal component of \PV,
\begin{equation}
\label{eq:conditionOnPV_ii}
\frac\partial{\partial\x_\alpha}\intD\,\AV\cdot\Rfrak\PV=q_\alpha \AV(\x_\alpha),
\end{equation}
would make the first term of \(H'\) simplify. However, it is not known whether the conditions \eref{eq:conditionOnPV} and \eref{eq:conditionOnPV_ii} can be simultaneously met in general. Nevertheless, we show that in the special case of the electric-dipole approximation to be performed in the next step, both conditions can be satisfied.

At this point, we summarize that under the condition \eref{eq:conditionOnPV_ii}, the Hamiltonian would have the form
\begin{equation}
\label{eq:HamiltonianPZW_ii}
H'=\sum_\alpha\frac{{\p'_\alpha}^2}{2m_\alpha}+\frac1{2\epsNot}\intD\,\PV^2-\frac1\epsNot\intD\,\DV\cdot\PV+H_\text{field}',
\end{equation}
where the kinetic term manifests the coincidence of the canonical momentum \(\p'_\alpha\) with the kinetic momentum of particle \(\alpha\), eliminating problem (i) listed after \Eref{eq:H_CG:ED}. We introduced the displacement field \(\DV\equiv\epsNot\EV+\PV\), about which, given that \(\PiV=\epsNot\partial_t\AV=-\Rfrak\EV\), it holds that \(\PiV'=-\Rfrak\DV=-\DV.\) The second equality holds because of \Eref{eq:polarizationField} and Gauss's law. \(H_\text{field}'\) is formally equivalent to \(H_\text{field},\) only with the transformed field momentum instead of the Coulomb-gauge one.

We now move from the description of point charges towards that of atoms in this picture. The polarization field is \(\sum_A\PV_A\), and since the atoms are spatially separated,
\begin{equation}
\intD\,\PV^2=\sum_A\intD\,\PV_A^2,
\end{equation}
therefore the first two terms of Hamiltonian \eref{eq:HamiltonianPZW_ii} give the internal energy of the atoms. In the electric-dipole approximation of atoms
\begin{equation}
\PV_A(\x)=\lp\sum_{\alpha\in A}q_\alpha\x_\alpha\rp\longwl\delta(\x-\x_A)\equiv\dv_A\,\longwl\delta(\x-\x_A),
\end{equation}
\(\dv_A\) being the electric dipole moment of atom \(A\). The function \(\longwl\delta\) behaves as a delta function over a spatial scale that is larger than the size of the atoms, while on the intra-atomic scale it is \emph{defined} such that condition \eref{eq:polarizationField} be satisfied (clearly, for a nonzero dipole moment, the charges cannot be at exactly the same position). With this definition, condition \eref{eq:conditionOnPV_ii} is met under our assumption that \(A(\x_\alpha)=A(\x_A)\). 

With the two conditions being satisfied, we can proceed from Hamiltonian \eref{eq:HamiltonianPZW_ii} to obtain the electric-dipole Hamiltonian in this picture:
\begin{subequations}
\label{eq:HamiltonianED}
\begin{equation}
H'_\text{ED}=\sum_A\lp H_A'-\dv_A\cdot\frac{\DV(\x_A)}\epsNot\rp+H_\text{field}',
\end{equation}
where the single-atom Hamiltonian has the form:
\begin{equation}
\label{eq:SingleAtomNew}
H_A'=\sum_{\alpha\in{A}}\frac{{\p'_\alpha}^2}{2m_\alpha}+\int_{\text{supp}(\PV_A)}d^3r\,\PV_A^2.
\end{equation}
\end{subequations}
In the second term, the domain of the integration can be restricted to the support of \(\PV_A\), so that unless the atom is very close to any of the boundary surfaces, the single-atom Hamiltonian is not at all affected by the presence of the boundaries. The intra-atomic Coulomb term (equivalent to the second term of the Hamiltonian \eref{eq:SingleAtomCG}) can be recovered from this same term, whereupon the remainder gives what is usually termed the dipole self-energy in this picture. This, however, does not concern us here because our agenda is to define the atomic levels \emph{in this picture} simply on the basis of the full single-atom Hamiltonian \eref{eq:SingleAtomNew}. For all practical purposes, the description of atoms is restricted to a few selected discrete energy levels, which can be taken phenomenologically from spectroscopic data. We note that the ``atom'' is not a gauge-invariant concept. The phenomenological replacement of the atom with a simple level structure (two-level, lambda, etc.) can be safely performed in the gauge of the new Hamiltonian \eref{eq:HamiltonianED}, because it is free from the problems listed above. Here, (i) the canonical momentum coincides with the kinetic one, (ii) the awkward A-square term has disappeared, as have (iii) the two Coulomb terms, describing atom-atom and atom-boundary interaction. In \(H'_\text{ED}\), the boundary enters only via the displacement field \DV, hence the atoms interact only via the retarded radiation field.

For quantizing the theory, we introduce the transverse modes as solutions to the constraint vectorial Helmholtz equation \footnote{It can be proven that the set of the transverse modes, that is, the eigenvectors corresponding to non-negative eigenvalues span \(\ker{\text{div}_0}\), and that the subspace of zero-frequency modes coincides with \CHgroup, that is, \(\omega_\lambda=0\) if and only if \(\modefunction\in\CHgroup\). Hence, on this degenerate finite dimensional subspace \CHgroup, an arbitrary basis can be chosen.}:
\begin{equation}
\label{eq:transverseModes}
\rot{\rot{\modefunction}}=\frac{\omega_\lambda^2}{c^2}\modefunction,\,\text{with}\,\div{\modefunction}=0\,\text{and}\,\at{\modefunction\times\n}\surfaceD=0.
\end{equation}
The vector potential \AV can be expanded in terms of these modes:
\begin{subequations}
\begin{equation}
\AV=\frac1\epsNot\sum_\lambda\lp\modefunction a_\lambda+\modefunctionconj a_\lambda^\dagger\rp,
\end{equation}
where \(a_\lambda\) is the annihilation operator of the corresponding mode, and this expansion was left invariant with respect to the Coulomb gauge. \DV is simply the canonical conjugate:
\begin{equation}
\DV=-\PiV'=i\epsNot\sum_\lambda\lp\modefunction a_\lambda-\modefunctionconj a_\lambda^\dagger\rp.
\end{equation}
\end{subequations}

We are now ready to systematically introduce the single-mode approximation, which is fundamental to the standard models of cavity QED (Dicke, Tavis--Cummings, Jaynes--Cummings). Our analysis has shown that even in the case of boundaries, when the possibility of a single-mode approximation arises at all, we still need the full mode expansion \eref{eq:transverseModes} for the cancellation of the A-square and the dipole-dipole interaction terms. Once this is done, in the new picture we can safely pick out one of the modes \modefunction. This is at variance with the approaches of Refs. \cite{Knight78,Keeling07}. For example, when the atoms can be treated as two-level systems, we obtain the Dicke model:
\begin{equation}
\label{eq:Dicke}
H_\text{Dicke}=\sum_A\lp\omega_A\,\sigma_z^{(A)}+g_A\lp a+a^\dagger\rp\sigma_x^{(A)}\rp+\omega\,a^\dagger a,
\end{equation}
where the three terms correspond one by one to the terms of the exact microscopic Hamiltonian \eref{eq:HamiltonianED} in the same order. We can thus conclude that these simplified models are better than generally expected.


This work was supported by the EU FP7 (ITN, CCQED-264666), the Hungarian National Office for Research and Technology under the contract ERC\_HU\_09 OPTOMECH, and the Hungarian Academy of Sciences (Lend\"ulet Program, LP2011-016). A. V. acknowledges support from the János Bolyai Research Scholarship of the Hungarian Academy of Sciences.

\bibliography{paper}

\begin{thebibliography}{11}
\expandafter\ifx\csname natexlab\endcsname\relax\def\natexlab#1{#1}\fi
\expandafter\ifx\csname bibnamefont\endcsname\relax
  \def\bibnamefont#1{#1}\fi
\expandafter\ifx\csname bibfnamefont\endcsname\relax
  \def\bibfnamefont#1{#1}\fi
\expandafter\ifx\csname citenamefont\endcsname\relax
  \def\citenamefont#1{#1}\fi
\expandafter\ifx\csname url\endcsname\relax
  \def\url#1{\texttt{#1}}\fi
\expandafter\ifx\csname urlprefix\endcsname\relax\def\urlprefix{URL }\fi
\providecommand{\bibinfo}[2]{#2}
\providecommand{\eprint}[2][]{\url{#2}}

\bibitem[{\citenamefont{Rza\ifmmode~\dot{z}\else \.{z}\fi{}ewski
  et~al.}(1975)\citenamefont{Rza\ifmmode~\dot{z}\else \.{z}\fi{}ewski,
  W\'odkiewicz, and \ifmmode~\dot{Z}\else \.{Z}\fi{}akowicz}}]{Rzazewski75}
\bibinfo{author}{\bibfnamefont{K.}~\bibnamefont{Rza\ifmmode~\dot{z}\else
  \.{z}\fi{}ewski}},
  \bibinfo{author}{\bibfnamefont{K.}~\bibnamefont{W\'odkiewicz}},
  \bibnamefont{and}
  \bibinfo{author}{\bibfnamefont{W.}~\bibnamefont{\ifmmode~\dot{Z}\else
  \.{Z}\fi{}akowicz}}, \bibinfo{journal}{Phys. Rev. Lett.}
  \textbf{\bibinfo{volume}{35}}, \bibinfo{pages}{432} (\bibinfo{year}{1975}).

\bibitem[{\citenamefont{Knight et~al.}(1978)\citenamefont{Knight, Aharonov, and
  Hsieh}}]{Knight78}
\bibinfo{author}{\bibfnamefont{J.~M.} \bibnamefont{Knight}},
  \bibinfo{author}{\bibfnamefont{Y.}~\bibnamefont{Aharonov}}, \bibnamefont{and}
  \bibinfo{author}{\bibfnamefont{G.~T.~C.} \bibnamefont{Hsieh}},
  \bibinfo{journal}{Phys. Rev. A} \textbf{\bibinfo{volume}{17}},
  \bibinfo{pages}{1454} (\bibinfo{year}{1978}).

\bibitem[{\citenamefont{Rza\ifmmode~\mbox{\c{}}\else \c{}\fi{}zewski and
  W\'odkiewicz}(1991)}]{Rzazewski91}
\bibinfo{author}{\bibfnamefont{K.}~\bibnamefont{Rza\ifmmode~\mbox{\c{}}\else
  \c{}\fi{}zewski}} \bibnamefont{and}
  \bibinfo{author}{\bibfnamefont{K.}~\bibnamefont{W\'odkiewicz}},
  \bibinfo{journal}{Phys. Rev. A} \textbf{\bibinfo{volume}{43}},
  \bibinfo{pages}{593} (\bibinfo{year}{1991}).

\bibitem[{\citenamefont{Nataf and Ciuti}({2010})}]{Nataf10}
\bibinfo{author}{\bibfnamefont{P.}~\bibnamefont{Nataf}} \bibnamefont{and}
  \bibinfo{author}{\bibfnamefont{C.}~\bibnamefont{Ciuti}},
  \bibinfo{journal}{{Nature Comm.}} \textbf{\bibinfo{volume}{{1}}}
  (\bibinfo{year}{{2010}}).

\bibitem[{\citenamefont{Ciuti et~al.}(2005)\citenamefont{Ciuti, Bastard, and
  Carusotto}}]{Ciuti05}
\bibinfo{author}{\bibfnamefont{C.}~\bibnamefont{Ciuti}},
  \bibinfo{author}{\bibfnamefont{G.}~\bibnamefont{Bastard}}, \bibnamefont{and}
  \bibinfo{author}{\bibfnamefont{I.}~\bibnamefont{Carusotto}},
  \bibinfo{journal}{Phys. Rev. B} \textbf{\bibinfo{volume}{72}},
  \bibinfo{pages}{115303} (\bibinfo{year}{2005}).

\bibitem[{\citenamefont{Todorov et~al.}({2010})\citenamefont{Todorov, Andrews,
  Colombelli, De~Liberato, Ciuti, Klang, Strasser, and Sirtori}}]{Todorov10}
\bibinfo{author}{\bibfnamefont{Y.}~\bibnamefont{Todorov}},
  \bibinfo{author}{\bibfnamefont{A.~M.} \bibnamefont{Andrews}},
  \bibinfo{author}{\bibfnamefont{R.}~\bibnamefont{Colombelli}},
  \bibinfo{author}{\bibfnamefont{S.}~\bibnamefont{De~Liberato}},
  \bibinfo{author}{\bibfnamefont{C.}~\bibnamefont{Ciuti}},
  \bibinfo{author}{\bibfnamefont{P.}~\bibnamefont{Klang}},
  \bibinfo{author}{\bibfnamefont{G.}~\bibnamefont{Strasser}}, \bibnamefont{and}
  \bibinfo{author}{\bibfnamefont{C.}~\bibnamefont{Sirtori}},
  \bibinfo{journal}{{Phys. Rev. Lett.}} \textbf{\bibinfo{volume}{{105}}}
  (\bibinfo{year}{{2010}}).

\bibitem[{\citenamefont{Keeling}(2007)}]{Keeling07}
\bibinfo{author}{\bibfnamefont{J.}~\bibnamefont{Keeling}}, \bibinfo{journal}{J.
  Phys.: Cond. Mat.} \textbf{\bibinfo{volume}{19}}, \bibinfo{pages}{295213}
  (\bibinfo{year}{2007}).

\bibitem[{\citenamefont{Vukics and Domokos}(2012)}]{Vukics2012}
\bibinfo{author}{\bibfnamefont{A.}~\bibnamefont{Vukics}} \bibnamefont{and}
  \bibinfo{author}{\bibfnamefont{P.}~\bibnamefont{Domokos}},
  \bibinfo{journal}{Physical Review A - Atomic, Molecular, and Optical Physics}
  \textbf{\bibinfo{volume}{86}} (\bibinfo{year}{2012}).

\bibitem[{\citenamefont{Cohen-Tannoudji
  et~al.}(1997)\citenamefont{Cohen-Tannoudji, Dupont-Roc, and Grynberg}}]{CDG}
\bibinfo{author}{\bibfnamefont{C.}~\bibnamefont{Cohen-Tannoudji}},
  \bibinfo{author}{\bibfnamefont{J.}~\bibnamefont{Dupont-Roc}},
  \bibnamefont{and} \bibinfo{author}{\bibfnamefont{G.}~\bibnamefont{Grynberg}},
  \emph{\bibinfo{title}{Photons and Atoms}}
  (\bibinfo{publisher}{Wiley-Interscience}, \bibinfo{year}{1997}).

\bibitem[{\citenamefont{Dautray and Lions}(1990)}]{dautray}
\bibinfo{author}{\bibfnamefont{R.}~\bibnamefont{Dautray}} \bibnamefont{and}
  \bibinfo{author}{\bibfnamefont{J.-L.} \bibnamefont{Lions}},
  \emph{\bibinfo{title}{Mathematical Analysis and Numerical Methods for Science
  and Technology}}, vol.~\bibinfo{volume}{3} (\bibinfo{publisher}{Springer},
  \bibinfo{year}{1990}).

\bibitem[{\citenamefont{Binz and Alfred}(2010)}]{binz10}
\bibinfo{author}{\bibfnamefont{E.}~\bibnamefont{Binz}} \bibnamefont{and}
  \bibinfo{author}{\bibfnamefont{R.}~\bibnamefont{Alfred}},
  \bibinfo{journal}{Journal of Physics: Conference Series}
  \textbf{\bibinfo{volume}{237}}, \bibinfo{pages}{012006}
  (\bibinfo{year}{2010}).

\end{thebibliography}

\end{document}